\documentstyle[12pt,aasms4]{article}

\lefthead{}
\righthead{}

\def\asec {^{\prime\prime}}
\def\amin {^{\prime}}
\def\fsec  {{\rlap.}^{\rm s}}
\def\fasec {{\rlap.}^{\prime \prime}\hskip0.05em}
\def\deg  {^\circ}

\begin{document}

\title{THE PROBABLE DETECTION OF SN~1923A: \\
THE OLDEST RADIO SUPERNOVA?}

\author{Christopher R. Eck\altaffilmark{1},
Douglas A. Roberts\altaffilmark{2},
John J. Cowan\altaffilmark{1}
and David Branch\altaffilmark{1}}

\begin {center}
eck@mail.nhn.ou.edu, dougr@ncsa.uiuc.edu, cowan@mail.nhn.ou.edu,
branch@mail.nhn.ou.edu
\end {center}

\altaffiltext{1}{Department of Physics and Astronomy,
University of Oklahoma, Norman, OK 73019}

\altaffiltext{2}{Department of Astronomy, University 
of Illinois, Urbana, IL 61801}

\begin{abstract}

Based upon the results of VLA observations, we report the detection of
two unresolved radio sources that are coincident with the 
reported optical position
of SN~1923A in M83.  For the source closest to the SN position, the flux
density was determined to be $0.30 \pm 0.05$ mJy at 20 cm and 
$0.093 \pm 0.028$ mJy at 6 cm.  The flux density of the second nearby 
source was determined to be $0.29 \pm 0.05$ at 20 cm and
$0.13 \pm 0.028$ at 6 cm.  Both sources are non-thermal with 
spectral indices of $\alpha = -1.0 \pm 0.30$ and $-0.69 \pm 0.24$, 
respectively.
SN 1923A has been designated as a Type II-P.  No Type II-P (other than
SN~1987A) has been detected previously in the radio.  
The radio emission from both sources appears to be fading with time.
At an age of approximately 68 years when we
observed it, this would be the oldest
radio supernova (of known age) yet detected.

\end{abstract}

\keywords{circumstellar matter --- supernovae: general 
--- supernovae: individual (SN~1923A) --- galaxies :
individual (NGC~5236)}

\section{Introduction}

Some Type II supernovae have exhibited radio emission within a few
years of explosion, including SNe 1970G in M101 (Gottesman et
al. 1972; Allen et al. 1976), 1978K in NGC 1313 (Ryder et al. 1993),
1979C in M100 (Weiler et al. 1986), 1980K in NGC 6946 (Weiler et
al. 1986, 1992), 1981K (van der Hulst et al. 1983; Weiler et
al. 1986), 1986J in NGC 891 (Rupen et al. 1987), 1988Z in MCG
+03-28-022 (Van Dyk et al. 1993b) and 1993J in NGC 3031 (Van Dyk et
al. 1994).  In addition SN 1987A was detected in the radio shortly
after outburst (see Turtle et al. 1987), but at a level significantly
below that seen from the more distant Type II supernovae and from the
Galactic SNR Cas A.  Radio emission was also detected for a brief
period from two Type Ibs, 1983N and 1984L (Sramek, Panagia, \& Weiler
1984; Weiler et al. 1986; Weiler \& Sramek 1988), and from the Type Ic
supernova 1990B (Van Dyk et al. 1993a).  No Type Ia SNe have been
detected in the radio despite several searches (e.g. \cite{eck95}).

Radio emission from supernova remnants (SNRs) is typically observed
long after the supernova phase.  Such factors as the density of the
local interstellar medium affect the turn-on time in models such as
those of Cowsik \& Sarkar (1984) based upon the Gull (1973) piston
model.  These models typically suggest a minimum of 100 years for the
formation and brightening of an SNR. SN~1987A is 
already indicating an
increase in brightness with time (Gaensler et al. 1997), but it is
entering a phase of delayed circumstellar interaction.

We (Cowan \& Branch 1982, 1985) have defined intermediate-age
supernovae, from $\sim$~10--300 years old, as spanning the period well
after the optical emission fades (typically about 2 years) and before
the turn-on of radio emission from an SNR (assumed to take at least
100 years).  This time period is critical in understanding the later
stages of stellar evolution. In particular, the circumstellar mass
loss rate for the supernova progenitors is a critical component in the
initiation and duration of radio emission in the models of Chevalier
(1982, 1984).

To study this transition period, we and others have attempted to
detect radio emission from intermediate-age supernovae. While there
have been a number of unsuccessful searches (see
\cite{eck95,eck96,eck98}), SNe 1950B \& 1957D in M83 (\cite {cow85,cow94}),
SN 1968D in NGC 6946 (Hyman et al. 1995) and
SN 1970G in M101 (\cite{cow91}) 
have been detected in the radio more than a decade after explosion. 
(SN 1961V
was also detected [\cite{bra85,cow88}], but there is uncertainty about
whether it really was a supernova [Filippenko et al. 1995].) We also
note that while the radio emission from SN~1980K has 
abruptly dropped after approximately ten years
(\cite{wei92,mon98}), SN~1979C (at a greater distance than SN~1980K) 
is still emitting at detectable levels
(\cite{wei91})

In this {\it Letter} we report the probable detection of SN~1923A in
M83, which at an age of $\sim$ 68 years would be the oldest
intermediate-age supernova of known age. 

\section{Observations and Data Reduction}

SN 1923A was discovered by C.\ Lampland (Lampland 1936) in M83,
classified as an 
SABc starburst galaxy at 4.1 Mpc (Saha et al. 1996) 
and home to 5 other supernovae 
(SNe 1945B, 1950B, 1957D, 1968L, 1983N).  Its position via 
offsets to the center of M83 is $109\asec$ E and $58\asec$ N.
A recent re-analysis of the map of M83 from CRB revealed two faint 
sources near the optical position of SN 1923A.

Observations of M83 were made at two epochs for each wavelength, 20
and 6 cm, at the Very Large Array (VLA)\footnote{The VLA is a
telescope of the National Radio Astronomy Observatory which is
operated by Associated Universities, Inc., under a cooperative
agreement with the National Science Foundation.}  in different
configurations such that the beam sizes were circular and approximately
the same for all observations.  The different ``hybrid''
configurations were used at the VLA to obtain circular beam sizes for
observations of M83, at low declination.  The phase and pointing
centers for all images were at R.A.(1950)$ = 13^{\rm h} 34^{\rm m}
12\fsec 2$, Dec.(1950)$ = -29\deg 35\amin 36\fasec 0$ and the strong
radio source 3C 286 was used as primary flux calibrator for all
observations.

Due to the non-thermal nature of the late-time radio emission from
supernovae, M83 was observed initially at 20 cm at the first epoch,
and then at 6 cm to determine the spectral index of the observed
sources.  At the second epoch a problem with the online systems at the
VLA corrupted the 20 cm data taken in the hybrid BnA
configuration. The observations were repeated later in B configuration
instead, resulting in a slightly different beam size than at the first
epoch at 20 cm.  Table 1 summarizes the relevant parameters for each
wavelength and epoch.

The data reduction was done using the Astronomical Image Processing
System (AIPS) software provided by the National Radio Astronomy
Observatory.  
Details of the original data reduction techniques are described in the 
papers regarding the observations of M83 (Cowan \& Branch 1985, CRB).
While a Maximum Entropy Method was used in the original image processing to 
reconstruct the diffuse structure, we have used CLEAN for the second 
epoch observations to 
get a higher signal-to-noise for isolated point sources (i.e.,  SN 1923A)
where the diffuse structure was not important.
We initially attempted to fit the two nearby point sources with two 
Gaussian components simultaneously using JMFIT, however, the relatively 
low signal-to-noise ratio of the two sources prevented convergence.
Therefore, all but one of the flux densities and positions were 
determined by fitting a quadratic function 
to each source, using the MAXFIT program of the AIPS package.
For the other datum, the fit using MAXFIT failed so that we report the peak 
flux using the AIPS routine IMSTAT.
A background level was estimated using TVSTAT at all wavelengths and 
epochs. The background level was determined to be on the order of the 
rms noise so it was included in the error estimation. 
PBCOR was used to correct the fluxes for primary beam attenuation. 
Images of the field of view can be seen in Figure 1 and results for
positions and fluxes are given in Table 2, along with the optical
supernova position from Pennington, Talbot, \& Dufour (1982).  
Figure 1a shows approximately half of M83 at 20 cm along with 
several sources CRB
observed, labeled here as in that paper.  Figure 1b shows the
region surrounding SN 1923A (zoomed in from Figure 1a) 
and includes both SN candidates.
Figure 1c is a 6cm map of approximately the same region as in 
Figure 1b.
To within the uncertainties\footnote{Without a fitting routine such as 
JMFIT to estimate positional uncertainty, we can only report the beamsize
as representative of the positional uncertainty although it is probably 
less than this.}, both unresolved sources are coincident 
with the optical 
position.  Although we tentatively report the western (closer) source 
as probably being from SN 1923A, we cannot exclude the possibility that 
either source may be SN 1923A.
Calculation of the spectral index, $\alpha$, ($S \propto
\nu^{\alpha}$) between 20 and 6 cm (at second epoch only) reveals that
both candidates are non-thermal, with the western source at $\alpha^{20}_{6} = 
-1.0 \pm 0.30$ and the eastern source at $\alpha^{20}_{6} = -0.69 \pm 0.24$.  
Some caution must be exercised in interpreting this value
since the flux densities were measured at slightly different times.
We can compare this value to the late-time spectral indices of other
Type II SNe ($\alpha^{20}_{6} = -0.57$ [1950B, CRB], $-0.23$
1957D, CRB]\footnote{Note: this value is at the first epoch
of observations since it is believed the SN has faded below the level
of an associated H II region at the second epoch (CRB)}, $-0.60$
[1961V, Cowan et al. 1988], 
$-0.92$ [1968D, Hyman et al. 1995], $-0.59$ [1970G, Cowan et
al. 1991], $-0.74$ [1979C, \cite{wei91}]).

Since there are 6 cm flux densities at two epochs, it is tempting to
try to fit the data to a power-law in time ($S \propto t^{\beta}$)
for comparison with other events ($\beta_{\rm 20cm} = -2.9, 
\beta_{\rm 6cm} = -1.7 [\mbox{1957D,CRB}]$,
$\beta_{\rm 20cm} = -1.95 [\mbox{1970G,\cite{cow91}}]$).
For SN 1923A we find $\beta_{\rm 6cm} =
-6.9 \pm 4.0$  and $-4.7 \pm 3.3$ for the western and eastern sources, 
respectively.  Since the flux densities appear to be 
decreasing with time, both sources 
are consistent with being in the later stages of radio supernova evolution.
While it is likely that both sources are 
fading radio SNe, it is not clear which source is SN 1923A.

\section{Discussion and Conclusions}

Although the detected sources are relatively weak, the non-thermal nature
and the apparent positional coincidence  
with the location of SN~1923A (as shown in Table 2)
make it probable that we have detected radio emission from this
supernova.  The sources are separated by about $3\fasec 5$ which, at the 
distance to M83 (4.1 Mpc), corresponds to almost 70 pc, thus the sources 
cannot both be from SN 1923A.  
Past optical studies of the site of SN~1923A have
shown evidence for an H~II region at or near the supernova
site. Rumstay \& Kaufman (1983) list an H II region --- no. 59 in
their paper --- within $1\asec-2\asec$ of the SN position as based upon
offsets with respect to the center of its parent galaxy.  Richter \&
Rosa (1984) refer to Rumstay \& Kaufman H II region no. 59 as an H II
region associated with and lying $1\asec$ from SN 1923A.  Pennington
et al. (1982) also note that SN 1923A appears to be
coincident with an H~II region.  A map of H II regions by de
Vaucouleurs, Pence, \& Davoust (1983), when overlaid with the scaled
radio map, has our source for SN~1923A directly over the H II region.
Since the progenitor star of SN~1923A has been estimated to have been
massive, $\simeq$ 18 M$_{\odot}$ (Pennington et al. 1982), it should
not have moved much prior to explosion, and the H~II region near the
radio source may be associated with the supernova's progenitor.  
The presence of an H~II region at the SN site reduces the chances that 
the two sources are actually background sources.  There
is evidence for other similar associations of radio supernovae (RSNe)
and H II regions (Van Dyk 1992), including SN~1957D 
(also estimated to have resulted
from a massive star) in M83 (CRB).  Clearly new examinations of the area
surrounding the site of SN~1923A in M83 to search for an optical
counterpart to the radio source are warranted.

On the basis of what is known about the shape of its light curve,
SN~1923A has been tentatively designated as a sub-luminous Type II-P
(Patat et al. 1994, Schaefer 1996). No Type II-P (other than SN~1987A)
has been detected previously in the radio.  In Figure 2, we plot the
radio luminosity at 20 cm of the new source assuming it to be
SN~1923A, along with several other extra-galactic RSNe and two Galactic
SNRs, as a function of time since outburst.  Data and fits (solid
lines) in Figure 2 for the well-studied Type II-L SN 1979C were taken
from Weiler et al. (1986, 1991), for the Type II-L SN 1980K from Weiler
et al. (1986, 1992), for the Type Ib SN 1983N from Weiler et al.
(1986) and Cowan \& Branch (1985), for the Type II-L SN 1970G from
Cowan et al. (1991), for the Type II-L SN 1968D from Hyman et al. (1995)
and for SNe 1950B \& 1957D from CRB.  The distance to M83 (4.1 Mpc) was 
taken to be the Cepheid--based distance to NGC
5253, (Saha et al. 1995), a fellow member of the Centaurus group.  As
Figure 2 illustrates, the luminosity of SN~1923A is comparable to, but
slightly below, the two other intermediate-age supernovae we have
detected in M83, SNe 1957D and 1950B.  (These two supernovae are
suspected to have had massive progenitors but their actual supernova
types are unknown.)  SN 1957D may actually be somewhat less luminous
than plotted, because emission from an associated H II region may have
contributed to the observed flux. The radio emission from SN~1923A
falls between that of Cas A and the Crab. 

At an age of approximately 68 years when last observed, SN~1923A
would be the oldest radio supernova yet discovered.  Detectable radio
emission from supernovae decades after explosion (but before the SNR
phase) may in fact be uncommon, as evidenced by the small class of
such known objects.  Observations of several other supernovae over a
number of years also support that conclusion.  While SN~1979C is still
detectable, SN~1980K has dropped off sharply after a decade of being
followed. Recently Montes et al. (1997) have reported early-time radio
emission from SN~1986E, while we were unable to detect this supernova
at an intermediate-age despite a deep VLA search (Eck et
al. 1996). Montes et al. (1997) argue that SN~1986E is a typical Type
II-L, similar to SN~1980K and the fading radio emission can be
adequately explained in terms of the Chevalier model.  

What is the cause of the radio emission of SN~1923A?  No Type II-P
events have been observed to undergo a prompt, bright circumstellar
interaction such as that of radio supernovae.  SN~1987A, having been a
sub-luminous Type II-P, underwent a prompt but dim circumstellar
interaction that would not have been detectable in a galaxy beyond the
local group, but now it is beginning what promises to be stronger,
delayed interaction with a detached circumstellar shell that
originated back in its red giant days.  If SN~1923A was a sub-luminous
Type II--P, it may now be fading from the kind of delayed interaction
that SN~1987A is just beginning.  

The key radio observations needed now (apart
from more firmly establishing the presence of a non-thermal radio
source at the site of SN~1923A) are to trace the radio evolution 
of these sources, one of which is likely to be SN 1923A, the 
oldest radio supernova yet detected.  While we noted above that theoretical 
models have suggested a minimum time of 100 years for the onset of the 
SNR (and a brightening) phase, radio emission from the supernovae at 
this age has never been previously detected or studied.  Our 
observations at one wavelength seem to indicate that the source closest to 
the SN position is still fading with time, but our uncertainties are large.
Additional observations of SN 1923A will help to understand more about the 
nature of radio emission as supernovae evolve from intermediate-age to 
the SNR phase.

\acknowledgements 
We wish to thank the anonymous referee for 
invaluable comments and suggestions to improve our paper. 
This work has been supported
in part by NSF grants AST-9314936, AST-9618332 and AST-9417102 at the
University of Oklahoma.  This paper was begun when J.J.C. was in
residence at the Institute for Theoretical Physics at the University
of California Santa Barbara and supported in part by the National
Science Foundation under Grant No. PHY94-07194.

\clearpage
\pagestyle{empty}

\noindent {\bf Figure Captions:}

\figcaption{
(a) A $1.515$ GHz (20 cm) contour map of approximately half 
of the field of view for M83 taken at the VLA in the 
B configuration on January 4, 1992.  The 
beam size ($\alpha \times \delta$) is $2\fasec 3 \times 5\fasec 2$ 
and the rms noise level is $0.050$ mJy beam$^{-1}$.  
The contour levels are at 
$0.17$, $0.27$, $0.60$, $0.89$, $1.19$, $1.49$, $1.79$, $2.08$, $2.38$, 
$2.68$, $2.98$, $5.95$, $8.93$, $11.9$, $14.9$, $17.9$, $20.8$, $23.8$, 
$26.8$, $29.5$ mJy beam$^{-1}$.  
SN 1957D (CRB) and SN 1923A are both visible 
in the map as well as several other sources from CRB including 
the radio bright central regions of M83. 
(b) A (20 cm) contour map of the region immediately surrounding the site 
of SN 1923A, magnified from the map in Figure 1a. 
Included in the map are both SN candidates
visible near the center of the map in Figure 1a.  A cross marks the 
optical position for SN 1923A from Pennington et al. (1982).  
The contour levels are $-0.071$, $0.071$, $0.14$, $0.17$, $0.19$, $0.21$, 
$0.24$, $0.26$, $0.27$, $0.29$, $0.30$ mJy beam$^{-1}$.  
The peak flux for the SN 1923A candidate is $0.30 \pm 0.05$ mJy beam$^{-1}$. 
(c) A (6cm) contour map of the region immediately surrounding the site 
of SN 1923A taken at the VLA in the CnB configuration on 
October 14, 1990.  The beamsize ($\alpha \times \delta$) is 
$3\fasec 5 \times 2\fasec 8$ and the rms noise level is $0.028$ 
mJy beam$^{-1}$.  A cross marks the
optical position for SN 1923A from Pennington et al. (1982). 
Both SN candidates are visible with peak 
flux positions consistent (to within uncertainties) with the 20 cm peak 
flux positions.  The angular separation of the peaks of the two sources 
is nearly identical ($3\fasec 5$) in the 6 cm map and in the 20 cm map.
The contour levels are $-0.065$, $0.065$, $0.076$, $0.087$, $0.098$, 
$0.11$, $0.12$, $0.13$ mJy beam$^{-1}$.  The peak flux for the candidate 
nearest the optical 
SN position is $0.093 \pm 0.03$ mJy beam$^{-1}$.
\label{fig1} }

\figcaption{
Radio luminosity  of SN 1923A compared to
several extra-galactic 
intermediate-age supernovae and Galactic SNRs 
at 20 cm as a function of time
since outburst. The peak flux densities for both sources are nearly 
equal, so we only plot a single data point as representative of 
the luminosity for SN 1923A.  
Data and fits (solid lines) for SN 1979C from Weiler et al. (1986, 1991),
SN 1980K from Weiler et al. (1986, 1992), 
SN 1983N from Weiler et al. (1986) and  Cowan \& Branch (1985), 
SN 1970G from Cowan et al. (1991), SN 1968D from Hyman et al. (1995) and
SNe 1950B \& 1957D from CRB. 
The dashed line is an extrapolation to the light curve of SN 1983N 
based on the Chevalier model. 
\label{fig2} }

\clearpage

\begin{table}

\begin{tabular}{rllcll}
\multicolumn{6}{c}{Parameters for Radio Observations}
\vspace{0.2cm} \\ \hline\hline
 & & & & & \\
 & \multicolumn{2}{c}{1983-1984 Observations} & & 
   \multicolumn{2}{c}{1990-1992 Observations} \\ [0.1cm]
 \cline{2-3} \cline{5-6} 
 & & & & & \\

Wavelength (cm) : & 20 & 6 & & 20 & 6 \\

Frequency (GHz) : & 1.446 & 4.873 & & 1.515 & 4.848 \\

Bandwidth per IF (MHz) :  &   12.5  &   25  & &   25  &   25   \\

Number of IFs :   &   2   &   2   & &   2   &   2 \\

Observation dates : & 15 Dec 1983 & 13 Mar 1984 & & 4 Jan 1992 & 14 Oct 1990
\\

Observing time (hr) : & 6 & 6.5 & & 8 & 8.5 \\

VLA Configuration : & BnA & CnB & & B & CnB \\

Primary beam HPBW : & $30\amin$ & $8\amin$ & & $30\amin$ & $8\amin$ \\

Resolution ($\alpha \times \delta$) : & $3\fasec 5 \times 3\fasec 5$ & 
$3\fasec 9 \times 2\fasec 8$ & & $2\fasec 3 \times 5\fasec 2$ & 
$3\fasec 2 \times 2\fasec 8$ \\

Rms noise (mJy beam$^{-1}$) : & $0.19$ & $0.054$ & & $0.050$ & $0.028$ \\

 & & & & & \\ \hline

\end{tabular}

\end{table}

\clearpage

\begin{table}

\begin{tabular}{lllccc}
\multicolumn{6}{c}{Radio Observations} 
\vspace{0.2cm} \\ \hline\hline
 & & & & \\

 & & & \multicolumn{2}{c}{Peak Flux (mJy)} & \\ \cline{4-5}

\multicolumn{1}{c}{Source} & \multicolumn{1}{c}{R.A.(1950)$^{\rm b}$}
 & \multicolumn{1}{c}{Dec.(1950)$^{\rm c}$} & 
   \multicolumn{1}{c}{20 cm} & \multicolumn{1}{c}{6 cm} & 
   \multicolumn{1}{c}{Age (yrs)} 
\vspace{0.2cm} \\ \hline

SN 1923A?$^{\rm a}$ & $13^{\rm h}34^{\rm m}19\fsec 92$ &
$-29\deg 35\amin 46\fasec 3$ &  
  $< 0.57^{\rm d}$ & $\cdots$ & $60.6$ \\
 & & & $\cdots$ & $0.19 \pm 0.05$ & $60.8$ \\

 & & & $\cdots$ & $0.093 \pm 0.03$ & $67.4$ \\
 & & & $0.30 \pm 0.05$ & $\cdots$ & $68.6$ \vspace{0.3cm} \\ 

Eastern Source & $13^{\rm h}34^{\rm m}20\fsec 18$ & 
$-29\deg 35\amin 46\fasec 6$ & $<0.57^{\rm d}$ & $\cdots$ & $60.6$ \\
  & & & $\cdots$ & $0.21 \pm 0.05$ & $60.8$ \\

   & & & $\cdots$ & $0.13 \pm 0.03$ & $67.4$ \\
   & & & $0.29 \pm 0.05$ & $\cdots$ & $68.6$ \\ 
\hline 

\end{tabular}

$^{\rm a}$ Optical position at 
$\alpha(1950) = 13^{\rm h}34^{\rm m}20^{\rm s} \pm 0\fsec 02$, 
$\delta(1950) = -29\deg 35\amin 48\asec \pm 0\fasec 24$ from 
Pennington, Talbot, \& Dufour (1982).  \newline
$^{\rm b} \pm 0\fsec 18$ \newline
$^{\rm c} \pm 5\fasec 2$ \newline
$^{\rm d} 3\sigma$ upper limit

\end{table} 

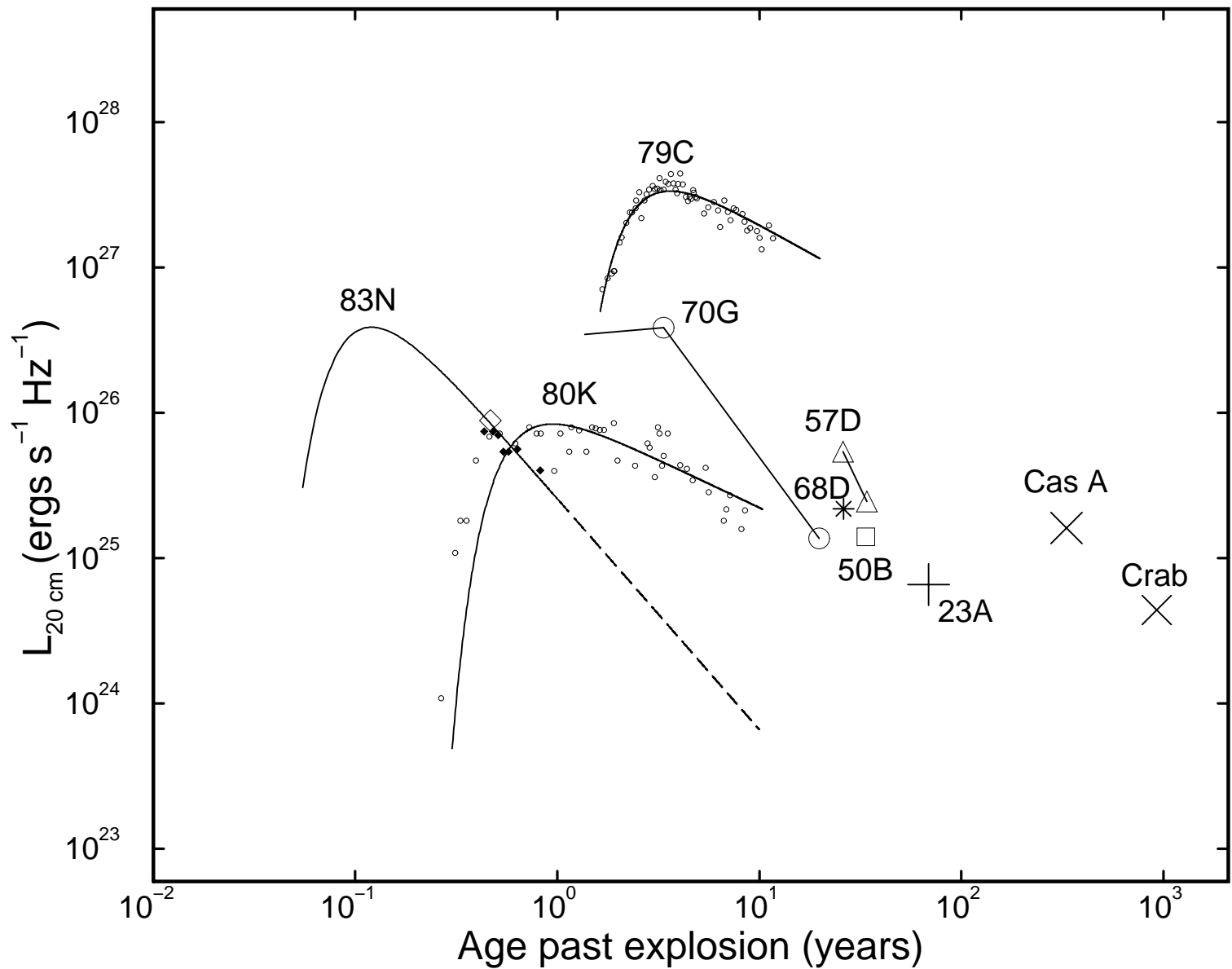
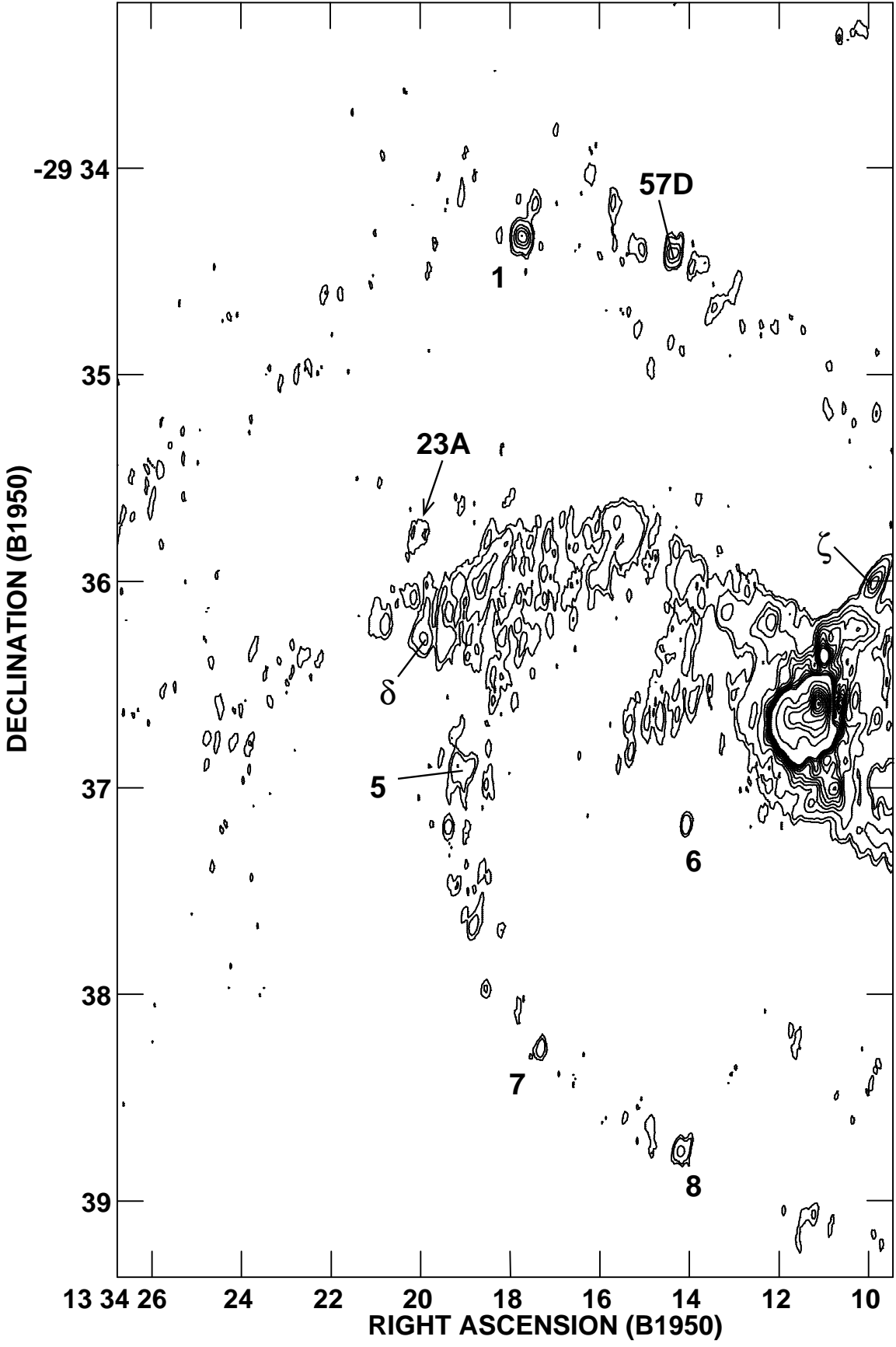
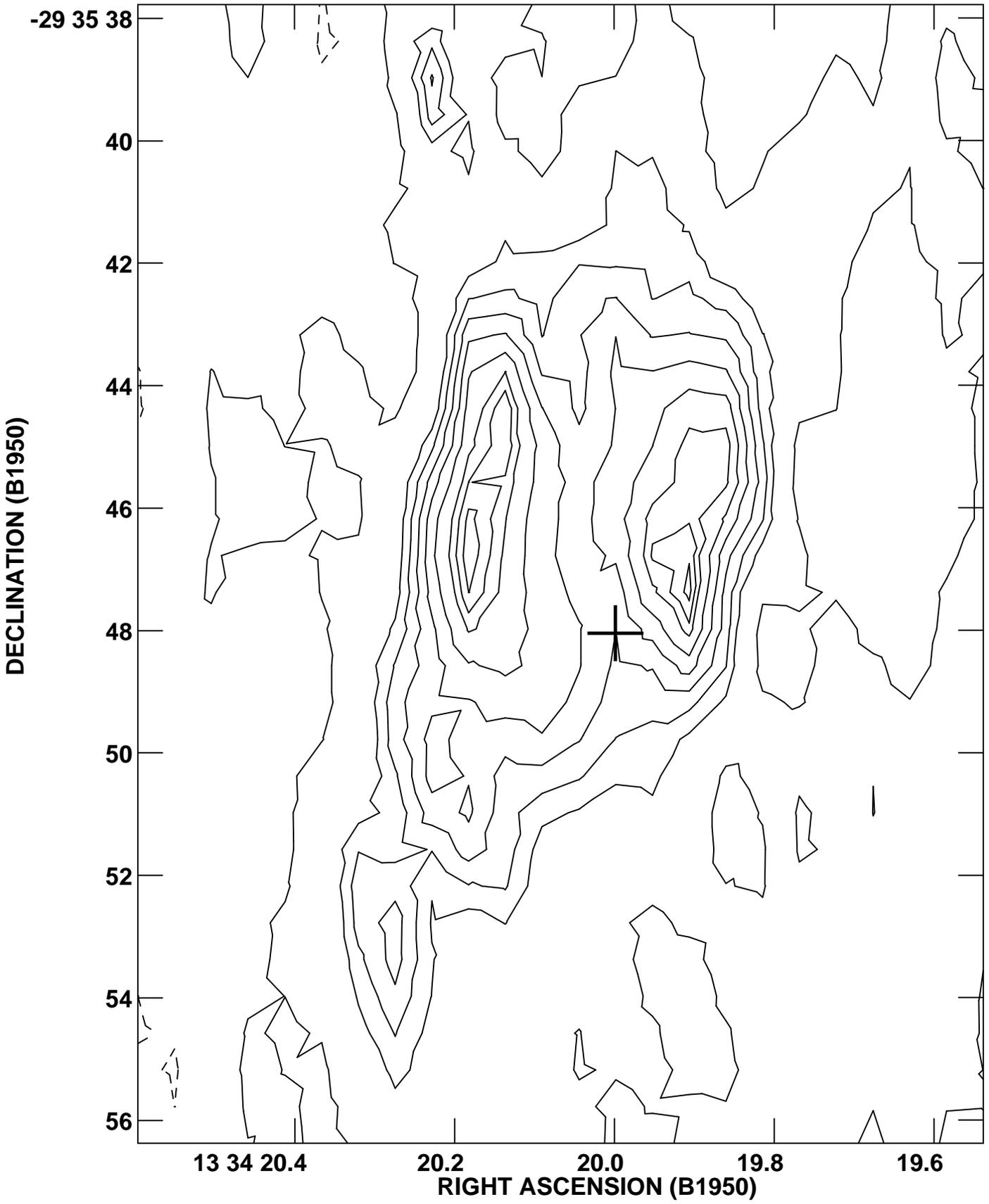
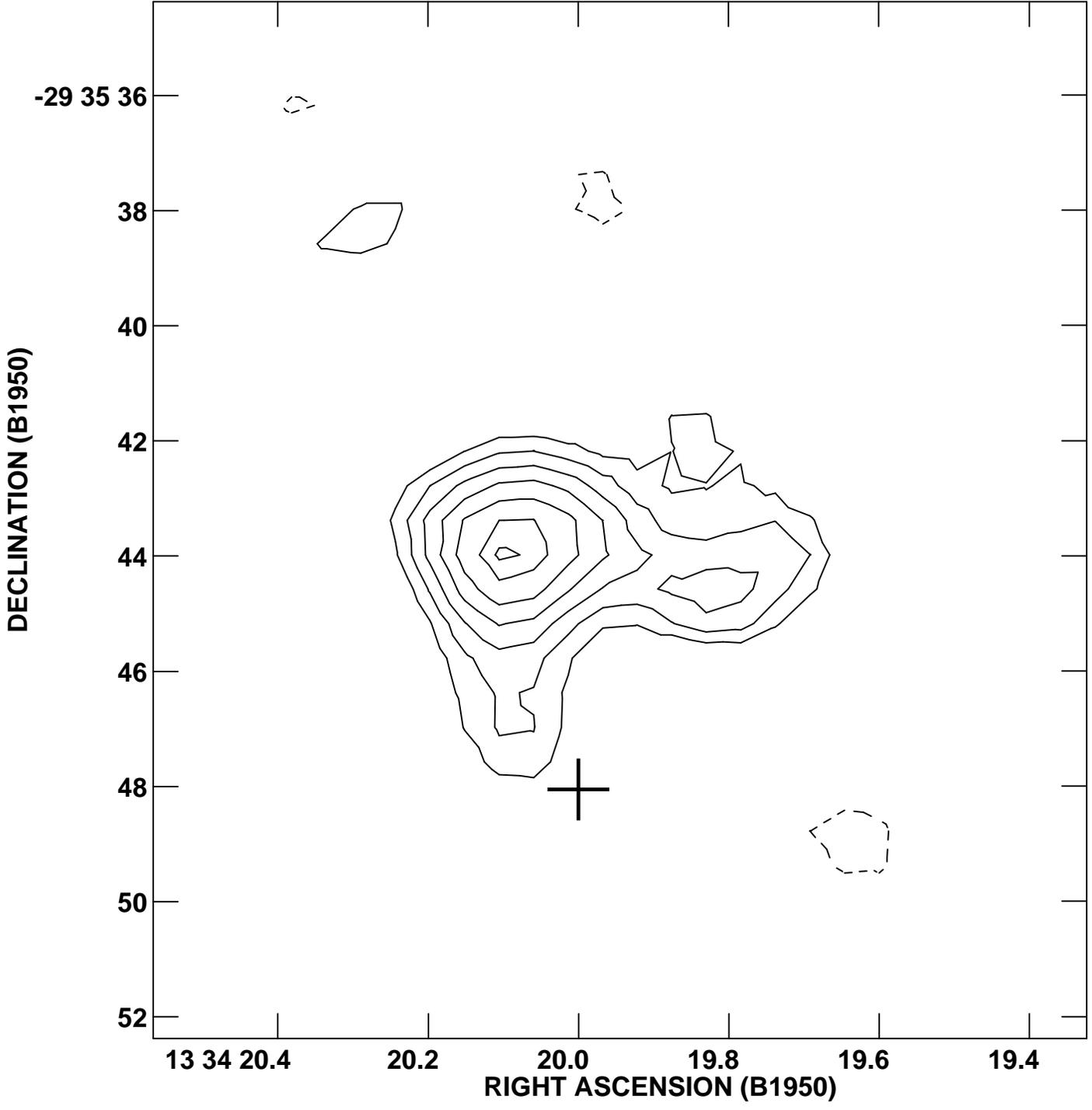
\end{document}